\documentclass[11pt]{amsart}
\usepackage{amsmath,amssymb}

\def\({\left(}  \def\){\right)}
\def\[{\left[}  \def\]{\right]}
\def\OL{\overline}
\def\UL{\underline}
\def \T{\ldots}
\def\W{\widetilde}
\def \RR{{\rm {\mathbb R}}}
\def \NN{{\rm {\mathbb N}}}
\def \ZZ{{\rm {\mathbb Z}}}

\newcommand{\mc}{\mathcal}
\newtheorem{theorem}{Theorem}
\newtheorem{lemma}{Lemma}

\newtheorem {corollary}{Corollary}
\newtheorem{prevtheorem}{Theorem}

\newcommand{\beq}{\begin{equation}}
\newcommand{\eeq}{\end{equation}}

\begin{document}

\title[Discrete Weighted Minimal Riesz Energy] {Asymptotics for Discrete Weighted Minimal Riesz Energy Problems on Rectifiable Sets}
\author [S.V. Borodachov]  {S.V. Borodachov$^{1}$}
\address{Center for Constructive Approximation, Department of Mathematics, \hspace*{.1in}
Vanderbilt University,
Nashville, TN 37240, USA  }

\email {sergiy.v.borodachov@vanderbilt.edu}
\author[D.P. Hardin]{D.P. Hardin$^{2}$}
 \email{doug.hardin@vanderbilt.edu}
\author[E. B. Saff]{E. B. Saff$^{3}$}
\email{edward.b.saff@vanderbilt.edu}

\thanks{\\[-.15in] $^{1}$Research of this author was conducted as a graduate student under the supervision of E.B. Saff and D. P. Hardin at Vanderbilt University. \\ \noindent $^{2}$The research of this author  was supported, in part,
by the U. S. National Science Foundation under grants DMS-0505756 and DMS-0532154. \\ \noindent $^{3}$The research of this author  was supported, in part,
by the U. S. National Science Foundation under grant DMS-0532154.}

\keywords{
Minimal discrete Riesz energy, Best-packing, Hausdorff measure, Rectifiable sets, non-Uniform distribution of points, Power law potential, Separation radius}
\subjclass{Primary 11K41, 70F10, 28A78; Secondary 78A30, 52A40}

\begin{abstract}
Given a compact $d$-rectifiable set $A$ embedded in Euclidean
space and a distribution $\rho(x)$ with respect to $d$-dimensional
Hausdorff measure on $A$, we address the following question: how
can one generate optimal configurations of $N$ points on $A$ that
are ``well-separated" and have asymptotic distribution $\rho (x)$
as $N\to \infty$? For this purpose we investigate minimal weighted
Riesz energy points, that is, points interacting via the weighted power law
potential $V=w(x,y)\left|x-y\right|^{-s}$, where $s>0$ is a fixed
parameter and $w$ is suitably chosen. In the unweighted case
($w\equiv 1$) such points for $N$ fixed tend to the solution of
the best-packing problem on $A$ as the parameter $s\to \infty$.
     \end{abstract}

\maketitle


\section {Introduction.}
Points on a compact set $A$ that minimize certain energy functions often have desirable properties that reflect 
special features of $A$.  For $A=S^2$, the unit sphere in $\RR^3$, the determination of minimum Coulomb
energy points is the classic problem of Thomson \cite{Mel77, Bow00}.  Other energy functions on higher dimensional 
spheres give rise to equilibrium points that are useful for a variety of applications including coding theory \cite{ConSlo99}, cubature formulas \cite{SloWom}, and  
the generation of finite normalized  tight frames \cite{BenFic03}.  In this paper, we shall consider a generalized Thomson problem, namely minimum energy points for weighted Riesz potentials on rectifiable sets.  Our focus is on the hypersingular
case when short range interactions between points is the dominant effect.  Such energy functions are not 
treatable with classical potential theoretic methods, and so require different techniques of analysis.  

 Let $A$ be a compact set in $\RR^{d'}$ whose $d$-dimensional Hausdorff measure,
$\mathcal{H}_d(A)$, is finite.
For  a collection of $N(\ge 2)$ distinct points
$\omega_N:=\{x_1,\T,x_N\}\subset A$, a non-negative weight function $w$ on $A\times A$ (we shall specify additional conditions on $w$ shortly),
and $s>0$,
 the  {\em weighted Riesz $s$-energy of $\omega_N$} is defined by
$$
E^w_s(\omega_N):=\sum_{1\leq i\neq j\leq N}{\frac
{w(x_i,x_j)}{\left |x_i-x_j\right|^s}}=\sum_{i=1}^N\sum_{ {j=1} \atop { j\neq i}}^N\frac{w(x_i,x_j)}{\left|x_i-x_j\right|^s},
$$
while  the {\em $N$-point
weighted Riesz $s$-energy of $A$} is defined by
\beq \label{c2'}\mathcal E^w_s(A,N):=\inf \{E^w_s(\omega_N) : \omega_N\subset A ,
 |\omega_N |=N\},
\eeq where $|X|$ denotes
the cardinality of a set $X$.  Since, for the weight $\W w(x,y) :=(w(x,y)+w(y,x))/2$, we have
$$E^w_s(\omega_N)=E^{\W w}_s(\omega_N)=2\sum\limits_{1\leq i< j\leq N}{\frac
{\W w(x_i,x_j)}{\left|x_i-x_j\right|^s}}, $$
 we shall assume, without loss of generality, throughout this paper that $w$ is symmetric, i.e.,
 $w(x,y)=w(y,x)$ for $x,y\in A$.
We call $w:A\times A\to [0,\infty]$ a  {\em CPD-weight function on $A\times A$} if
\begin{enumerate}
\item[(a)] $w$ is continuous (as a function on $A\times A$) at $\mathcal{H}_d$-almost every point of
the diagonal $D(A):=\{(x,x) : x\in A\}$,
\item[(b)] there is some neighborhood $G$ of $D(A)$ (relative to $A\times A$) such that $\inf_G w>0$,  and
\item[(c)] $w$ is bounded on any closed subset $B\subset A\times A$ such that $B\cap D(A)=\emptyset$.
\end{enumerate}
Here CPD stands for (almost) continuous and positive on the diagonal.
In particular, conditions (a), (b), and (c) hold if $w$ is bounded on $A\times A$ and continuous and positive
at every point  of the diagonal  $D(A)$ (where
continuity at a diagonal point $(x_0,x_0)$ is meant in the sense of limits taken on
$A\times A$).

If $w\equiv 1$ on $A\times A$ (which we refer to as the {\em unweighted} case),  we write
$E_s(\omega_N)$ and $\mathcal E_s(A,N)$ for $ E^w_s(\omega_N)$ and $\mathcal E^w_s(A,N)$, respectively.   For the trivial cases $N=0$ or $1$ we put
$E_s(\omega_N)=\mathcal E_s(A,N)=E^w_s(\omega_N)=\mathcal
E^w_s(A,N)=0$.

We are interested in the geometrical properties of optimal
$s$-energy $N$-point configurations for a set $A$; that is, sets
$\omega_N$ for which the infimum in (\ref{c2'}) is attained.
Indeed, these configurations are useful in statistical sampling,
weighted quadrature, and computer-aided geometric design where the
selection of a ``good''  finite (but possibly large) collection of
points is required to represent a set or manifold $A$. Since the
exact determination of optimal configurations seems, except in a
handful of cases (cf.
\cite{Yud93,KolYud94,And96,KolYud97,And97,DraLegTow02}), beyond
the realm of possibility, our focus is on the asymptotics of such
configurations.
Specifically, we consider the following questions.\\

(i)  What is the asymptotic behavior of the quantity $\mathcal E^w_s(A,N)$ as $N$ gets large?

(ii)  How are optimal point
configurations distributed as  $N\to \infty$?

(iii) What estimates can be given for the minimal pairwise distance
between points in optimal configurations for large $N$?\\




In the unweighted case, much is known regarding these questions.
In particular, when
$s< {\rm dim}\ A$ (the Hausdorff dimension of  $A$),  the  limit distribution of   optimal
$N$-point configurations is given by the {\em equilibrium measure} $\lambda_{s,A}$
that minimizes the continuous
energy integral
$$
I_s(\mu):= \iint \limits_{A\times A}{\frac {1}{\left|x-y\right|^s}\ d\mu (x)\ \! d\mu (y)}
$$
over the class $\mathcal{M}(A)$ of (Radon) probability measures $\mu$ supported on $A$.  In addition,  the asymptotic order of the Riesz $s$-energy is $N^2$;   more precisely we have  $\mathcal E_s(A,N)/N^2\to I_s(\lambda_{s,A})$ as $N\to \infty$   (cf. \cite[Section II.3.12]{LanFMPT}).
 In the case when $A=S^d$, the unit sphere in $\RR^{d+1}$, the equilibrium measure is simply the normalized surface area measure.

  If $s\geq {\rm dim }\ A$, then $I_s(\mu)=\infty$ for every $\mu\in \mathcal{M}(A)$ and potential
  theoretic methods cannot be used. However, by using techniques from geometric measure theory, it was recently
  shown in    \cite{HarSaf04} that
  when $A$ is a  $d$-rectifiable manifold  of positive $d$-dimensional Hausdorff measure and $s\ge d$,
   optimal $N$-point configurations are uniformly distributed (as $N\to\infty$) on $A$ with respect to $d$-dimensional Hausdorff measure restricted to $A$.  The assertion for the case $s=d$ further requires that $A$ be a subset of a $C^1$ manifold (see Theorem~\ref{thmA} in Section 1.1).

Our motivation for considering the {\it weighted} minimal Riesz energy problem
is for the purpose of obtaining point sets that are distributed according to a specified non-uniform
density such as might be used as nodes for weighted integration or   in computer
modeling of surfaces where more points are needed in  regions with higher curvature.
 In this paper we shall show that for a compact  $d$-rectifiable set  $A$ having positive $d$-dimensional Hausdorff measure,
  $N$-point configurations for $A$ minimizing the weighted Riesz $s$-energy
are distributed asymptotically with density proportional to $(w(x,x))^{-d/s}$ provided $s\ge d$. (This continues
 to be true even when $w$ has a finite number of zeros on the diagonal, provided their order is less than $s$.)
Even in the unweighted case, these results extend those of \cite{HarSaf04} obtained for the class of $d$-rectifiable manifolds to the more general class of  $d$-rectifiable sets.

 For the remainder of this introduction we provide the needed notation and discuss known results.  Section 2 is 
 devoted to the statements of the main results of this paper.   The detailed proofs of these main results, which utilize
 the basic lemmas described in Section 3, are provided in Sections 4, 5 and 6.  



\subsection{Notation and previous results.}

It is helpful to keep in mind that minimal discrete $s$-energy
problems can be considered as a bridge between   logarithmic
energy problems and best-packing ones.  Indeed, in the unweighted
case ($w(x,y)\equiv 1$) when $s\to 0$ and $N$ is fixed, the
minimal energy problem turns into the problem for the logarithmic
potential energy
$$
\sum\limits_{1\leq i\neq j\leq N}{\log \frac
{1}{\left|x_i-x_j\right|}},
$$
which is minimized over all $N$-point configurations
$\{x_1,\ldots,x_N\}\subset A$. This problem is equivalent
to the  maximization of the product
$$
\prod_{1\leq i\neq j\leq N}{\left|x_i-x_j\right|}.
$$
For planar sets, such optimal points are known as  {\it Fekete points}.         
(For the case when $A=S^2$, the polynomial time generation of  ``nearly optimal'' points for the logarithmic
energy is the focus of one of S. Smale's ``problems for the next century''; see \cite{S}.)

On the other hand,
when $s\to\infty$, and $N$ is fixed in the unweighted case, we get the best-packing
problem (cf. \cite{LF-T}, \cite{ConSlo99}); i.e.,  the problem of finding $N$-point configurations
$\omega_N\subset A$ with the largest separation radius:
\begin {equation}\label {sepr}
\delta (\omega_N):=\min\limits_{1\leq i\neq j\leq
N}{\left|x_i-x_j\right|}.
\end {equation}

In this paper we will consider the case $s\geq {\rm dim}\ A$.
Let $\mathcal L_{d'}$ be the Lebesgue measure in
$\RR^{d'}$
and
$\mathcal H_d$ be the $d$-dimensional Hausdorff measure in $\RR
^{d'}$ normalized so that its restriction to $\RR^d \subset
\RR^{d'}$ is $\mathcal L_d$.
Denote by $B_{d'}(x_0,r)$ 
the open ball in $\RR^{d'}$
centered at the point $x_0$ with radius $r>0$ and set
\begin{equation}\label{betad}
\beta _d:=\mathcal L_d(B_d(0,1))=\frac{2\pi^{d/2}}{d\ \Gamma(d/2)}.
\end{equation}
Given sequences
$\{a_N\}_{N=1}^\infty$ and $\{b_N\}_{N=1}^\infty$ of positive
numbers, we will write $a_N \sim b_N$, $N\to \infty$, if
$\lim_{N\to \infty}{a_N/b_N}=1$.

Regarding questions (i) and (ii),
A.B.J. Kuijlaars and E.B. Saff \cite {KuiSaf98} proved  that for the unit sphere
$S^d$, there holds
$$
\mathcal E_d(S^d,N)\sim \frac{\beta_d}{\mathcal H_d(S^d)}N^2\log N,\
\ \ N\to \infty,
$$
and 
it is known that the distribution of the minimal energy points is
asymptotically uniform in this case.
For  one-dimensional rectifiable curves in $\RR^{d'}$, the paper by
A. Martinez-Finkelshtein et al \cite{MarMayRahSaf02}
provides answers in the unweighted case
to questions (i) and (ii), as well as question (iii) for {\em regular} Jordan
arcs or curves.

Question (iii) in the unweighted case has also been considered for several other
special cases.
B.E.J. Dahlberg \cite {Dal78} proved that if $A=S^d$ and $s=d-1$,
$d\geq 2$, or $A\subset \RR^3$ is a smooth surface and $s=1$,
$d=2$, there is a constant $C>0$ such that for every $s$-optimal
collection $\omega^\ast_N\subset A$ with $N$ points
\begin{equation}\label{deltabnd}
\delta (\omega^\ast_N)\geq CN^{-1/d}.
\end{equation}
In \cite{KuiSaf98} it was shown that
(\ref{deltabnd})
holds  for $A=S^d$ when $s>d$.

In   \cite{HarSaf04} and \cite{HarSaf04N}, questions (i), (ii), and (iii) were addressed for a 
more general class of sets $A$
which we now describe.
First recall that a mapping $\phi:T\to \RR^{d'}$, $T\subset\RR^{d}$,  is  said to be a
     {\em Lipschitz mapping on $T$} if there is some constant $\lambda$ such that
     \begin{equation} \label{Lipdef}
   |\phi(x)-\phi(y)|\le \lambda |x-y|\qquad \text{for $x,y\in T$,}
    \eeq
    and that $\phi$ is said to be  a {\em bi-Lipschitz mapping on $T$
    (with constant $\lambda$)} if
     \beq \label{biLipdef}
   (1/\lambda)|x-y|\le|\phi(x)-\phi(y)|\le \lambda|x-y|\qquad \text{for $x,y\in T$}.
    \eeq
Following \cite{HarSaf04}, we say that a set $A\subset \RR^{d'}$ is a {\em $d$-rectifiable manifold}
if it is a compact subset of a finite union of bi-Lipschitz images of open sets in $\RR^d$.

 We further recall that if $A\subset\RR^{d'}$ is compact and $\nu$ and $\{\nu_N\}_{N=1}^{\infty}$
  are  Borel probability measures on $A$, then
  the sequence $\nu_N$ converges {\em weak-star} to $\nu$  (and we write $\nu_N {\ \ \ast \over}\!\!\!\!\to \nu$) if for any  function $f$ continuous on $A$,  we have
$$
\lim_{N\to\infty}{\int_{A}{fd\nu_N}}=\int_{A}{fd\nu}.
$$
Denote by $\delta _x$ the atomic probability measure in
$\RR^{d'}$ centered at the point $x\in \RR^{d'}$.

For future reference and comparison we now state some main results from \cite {HarSaf04}.

\begin {prevtheorem}\label{thmA}
Let $A\subset \RR^{d'}$ be a compact $d$-rectifiable manifold  and suppose     $s>d$. Then
\begin {equation}\label {s>d}
\lim_{N\to\infty}{\frac {\mathcal E_s(A,N)}{N^{1+s/d}}}=\frac { C_{s,d}}{\mathcal H_d(A)^{s/d}},
\end {equation}
where $C_{s,d}$ is a positive constant independent of $A$.

 Furthermore, if $\mathcal H_{d}(A)>0$, any asymptotically $s$-energy minimizing
 sequence of configurations
$\W\omega_N=\{x^N_1,\ldots,x^N_N\}$, $N=2,3,\ldots$,
 for $A$ is uniformly distributed with respect to $\mathcal{H}_d$; that is,
\begin {equation}\label {distr}
\frac 1N\sum\limits_{k=1}^{N}{\delta _{x^N_{k}}}{\ \ \ast
\over}\!\!\!\to \frac{\mathcal H_d|_A}{\mathcal H_d(A)},\ \ N\to\infty.
\end {equation}
\end{prevtheorem}

 By  {\em asymptotically $s$-energy minimizing} we mean
$$
E_s(\W\omega_N)\sim\mathcal E_s(A,N),\ \ N\to \infty .
$$

{\bf Remark.} For $s>d$, the constant $C_{s,d}$  appearing in (\ref{s>d}) of Theorem~\ref{thmA} can be represented  using the
energy for the unit cube in $\RR^d$ via formula (\ref{s>d}):
$$
C_{s,d}=\lim\limits_{N\to \infty}{\frac {\mathcal
E_s([0,1]^d,N)}{N^{1+s/d}}}, \ \ s>d.
$$
For $d=1$ and $s>1$, it was shown in  \cite{MarMayRahSaf02} that $C_{s,1}=2\zeta(s)$,
where $\zeta(s)$ is the classical Riemann zeta function.   However, for other values of $d$, the constant $C_{s,d}$ is as yet unknown.
For the case $d=2$, it is a consequence of results in \cite{KuiSaf98}  that
     \beq
     \label{Cs22}
C_{s,2}\le\left( \sqrt{3}/2\right)^{s/2}\zeta_{L}(s),
    \eeq
    where $\zeta_{L}(s)$ is the zeta function for the planar triangular
lattice $L$ consisting of points of the form $m(1,0)+n(1/2,\sqrt{3}/2)$ for $m,n\in\ZZ$.
It is conjectured in \cite{KuiSaf98} that in fact equality holds in (\ref{Cs22}).
Furthermore,  it is  shown in \cite{BorThesis} that as $s\to \infty$
$$\left(C_{s,2}\right)^{2/s}\to
\sqrt{3}/2,$$
which is consistent with this conjecture.

When $0<\mathcal{H}_{d}(A)<\infty$ we observe  that the minimum energy
experiences a transition in order of growth; namely,
as $s$ increases from values less than $d$ to values greater than $d$
the energy switches from order $N^{2}$ to order $N^{1+s/d}$ as $N\to \infty$.
As the following theorem from \cite{HarSaf04} describes, at the transition value
$s=d$,     the order of  growth is  $N^{2}\log N$.  In the proof of this fact,  a more  delicate analysis was utilized that
involved an additional regularity assumption on $A$.

\begin {prevtheorem}\label {thmB}
Let   $A$ be a compact subset of a $d$-dimensional
$C^1$-manifold in $\RR^{d'}$.  Then
\begin {equation}\label {s=d}
\lim_{N\to\infty}{\frac {\mathcal E_d(A,N)}{N^2\log N}}=\frac { \beta_d}{\mathcal H_d(A)},
\end {equation}
where $\beta_d$ is the volume of the $d$-dimensional unit ball  as defined in (\ref{betad}).

 Furthermore, if $\mathcal H_{d}(A)>0$, any asymptotically $d$-energy minimizing
 sequence of configurations
$\W\omega_N=\{x^N_1,\ldots,x^N_N\}$, $N=2,3,\ldots$,
 for $A$ is uniformly distributed with respect to $\mathcal{H}_d$; that is, (\ref{distr}) holds.
\end {prevtheorem}


Regarding separation results, the following was shown in \cite {HarSaf04}. If
$A\subset \RR^{d'}$ is a bi-Lipschitz image of a compact set from
$\RR^d$ of positive Lebesgue measure, then for every $s\geq d$
there is a constant $c_s>0$ such that
$$
\delta (\omega^\ast_N)\geq \begin {cases}  c_sN^{-1/d},  & s>d, \\ c_d(N\log N)^{-1/d}, & s=d, \end {cases}
$$
for every $s$-optimal $N$-point configuration $\omega^\ast_N$ on $A$.
This result was extended by S. Damelin and V. Maymeskul in
\cite {DamMay04} to a finite union of bi-Lipschitz
images of compact sets from $\RR^d$ .


\section{Main results.}\label {3}

In this paper we extend  Theorem A  to the class  of $d$-rectifiable sets; where, by a  {\em $d$-rectifiable
set $A\subset \RR^{d'}$},  we mean
the  image of a bounded set in $\RR ^d$ under a Lipschitz mapping   (cf.  \cite {FedGMT}).   Consequently,
  we relax the bi-Lipschitz
condition in Theorem A.  Furthermore,  both Theorems A and B are
extended to the case of weighted energy, and 
 separation estimates
for optimal configurations are obtained even when the
 Hausdorff dimension of the compact set is not necessarily integer.

Our first main result  extends  (\ref{s>d}) of Theorem A to the class of $d$-rectifiable sets.
\begin {theorem}\label {rectifiable}
Suppose $A\subset
\RR^{d'}$ is a closed $d$-rectifiable set. Then, for $s>d$,
\begin {equation}\label{w3.1}
\lim\limits_{N\to \infty}{\frac{\mathcal
E_s(A,N)}{N^{1+s/d}}}=\frac {C_{s,d}}{\mathcal H_d(A)^{s/d}},
\end{equation}
where $C_{s,d}$ is  as in Theorem~A.
\end{theorem}
A key component of this extension is a regularity result given in  Lemma~\ref{rl5} of Section 3
showing that   $\liminf_{N\to\infty} \mathcal E_s(K,N)/N^{1+s/d}$ is close to $\liminf_{N\to\infty} \mathcal E_s(A,N)/N^{1+s/d}$ for compact sets $K\subset A$  such that $\mathcal{H}_d(A\setminus K)$ is sufficiently small.

If $A$ is a compact set in $\RR^{d'}$ and $w$ is a CPD-weight function on $A\times A$, then for $s\geq d$ we define the {\em weighted Hausdorff measure } $\mathcal{H}_d^{s,w}$ on  Borel sets $B\subset A$ by
\begin{equation}
\mathcal{H}_d^{s,w}(B):=\int_{B}{(w(x,x))^{-d/s}d\mathcal H_d(x)},
\end{equation}
and its normalized form
\begin{equation}
 h_d^{s,w}(B):=\mathcal{H}_d^{s,w}(B)/\mathcal{H}_d^{s,w}(A).
\end{equation}

We say, that a sequence $\{\W\omega_N\}_{N=1}^\infty$ of $N$-point
configurations in $A$ is {\em asymptotically ($w,s$)-energy
minimizing for $A$} if
$$
E^w_s(\W\omega_N)\sim\mathcal
E^w_s(A,N),\ \ N\to\infty.
$$

The main results of this paper include  the following generalizations of Theorems~\ref{thmA}  and  \ref{thmB}.

\begin {theorem}\label{rth1}
Let $A\subset \RR^{d'}$ be a compact $d$-rectifiable set. Suppose 
$s>d$ and that $w$ is  a  CPD-weight
function on $A\times A$.  Then
\begin {equation}\label {w3j}
\lim\limits_{N\to \infty}{\frac{\mathcal
E^w_s(A,N)}{N^{1+s/d}}}=\frac{C_{s,d}}{\left[\mathcal{H}_d^{s,w}(A)\right]^{s/d}},
\end {equation}
where $C_{s,d}$ is as in Theorem~\ref{thmA}.

 Furthermore, if $\mathcal H_{d}(A)>0$, any asymptotically $(w,s)$-energy minimizing
 sequence of configurations
$\W\omega_N=\{x^N_1,\ldots,x^N_N\}$, $N=2,3,\ldots$,
 for $A$ is uniformly distributed with respect to $\mathcal{H}_d^{s,w}$;
that is,
\begin{equation} \label{wdistlim}
\frac 1N\sum\limits_{k=1}^{N}{\delta _{x^N_k}}{\ \ \ast \over}\!\!\!\to h_d^{s,w},\ \ N\to\infty.
\end{equation}
\end{theorem}

\begin {theorem}\label {wth1H}
Let   $A$ be a compact subset of a $d$-dimensional
$C^1$-manifold in $\RR^{d'}$ and suppose $w$  is  a CPD-weight function on
$A\times A$.  Then
\begin {equation}\label {q1}
\lim\limits_{N\to \infty}{\frac {\mathcal E^w_d(A,N)}{N^{2}\log
N}}=\frac{\beta_d}{\mathcal{H}_d^{d,w}(A)}.
\end {equation}

 Furthermore, if $\mathcal H_{d}(A)>0$, any asymptotically $(w,d)$-energy minimizing
 sequence of configurations
$\W\omega_N=\{x^N_1,\ldots,x^N_N\}$, $N=2,3,\ldots$,
 for $A$ is uniformly distributed with respect to $\mathcal{H}_d^{d,w}$; that is,
 (\ref{wdistlim}) holds with $s=d$.
\end {theorem}

{\bf Remarks.}
 In the case $\mathcal H_d(A)=0$, the right-hand sides
of (\ref {w3j}) and (\ref {q1}) are understood to be infinity.

Next we obtain   estimates for the
separation radius of optimal configurations on sets of arbitrary
 Hausdorff dimension $ \alpha$.   We remark that the normalization
 for the Hausdorff measure $\mathcal H_\alpha$ plays no essential role here.
\begin {theorem}\label {th2}
Let $0<\alpha\leq d'$.  Suppose $A\subset {\rm {\bf R}}^{d'}$ is a
compact set with $\mathcal H_\alpha (A)>0$ and let $w$ be a
 CPD-weight function   that is bounded and lower semi-continuous  on
$A\times A$. Then, for every $s\geq\alpha$ there is a constant
$c_s=c_s(A,w,\alpha)>0$ such that any $(w,s)$-energy minimizing
configuration $\omega^\ast_N:=\{x_{1,N},\ldots,x_{N,N}\}$ on $A$
satisfies the inequality
$$
\delta(\omega^\ast_N)=\min\limits_{1\leq i\neq
j\leq N}{\left|x_{i,N}-x_{j,N}\right|}\geq
\begin {cases}{c_sN^{-1/\alpha}, \ \ \ \ \ \ \ \ \  s>\alpha,}\cr
{c_\alpha(N\log N)^{-1/\alpha},\ s=\alpha, \ \ N\geq 2.}\end {cases}
$$
\end {theorem}


As a consequence of the proof of Theorem \ref {th2} we establish the following estimates.
Let $$\mathcal H^\infty_\alpha (A):=\inf\{\sum\limits_{i}{\({\rm diam}\ G_i\)^\alpha} : A\subset \bigcup_{i}G_i\}.$$

\begin{corollary} \label{cor1}Under the assumptions of Theorem \ref {th2},
 for $N\geq 2$,
$$
\mathcal E^w_s(A,N)\leq
\begin {cases}{ {M_{s,\alpha}\|w\|_{A\times A}}{\mathcal H^\infty_\alpha (A)^{-s/\alpha}}N^{1+s/\alpha}, \ \ \
s>\alpha,}\cr{M_\alpha N^2\log N,\ \ s=\alpha,}
\end{cases}
$$
where the constant $M_{s,\alpha}>0$ is independent of $A$, $w$ and $N$, and the constant $M_\alpha$ is independent of $N$.
\end{corollary}

In order to obtain a finite collection of points distributed with
a given density $\rho (x)$ on a $d$-rectifiable set $A$, we can take any $s> d$
and the weight \begin{equation} \label{wrhodef}
w(x,y):=(\rho (x)\rho (y)+|x-y|)^{-s/2d},
\end{equation} 
where the term $|x-y|$ is included to ensure that $w$ is locally bounded off of $D(A)$.  By Theorems
\ref {rth1} and \ref {wth1H} any asymptotically $(w,s)$-energy minimizing sequence of
$N$-point configurations will converge to the required distribution as
$N\to \infty$.  We thus obtain\\

\begin{corollary}\label{wrho}
Let $A\subset \RR^{d'}$ be a compact $d$-rectifiable set with $\mathcal{H}_d(A)>0$.
Suppose $\rho$ is a bounded  probability density on $A$ (with respect to $\mathcal H_d$) that is
  continuous $\mathcal H_d$-almost everywhere on $A$.  Then, for $s>d$ and $w$
  given by (\ref{wrhodef}), the normalized counting measures for  any asymptotically $(w,s)$-energy minimizing sequence of configurations $\omega_N$ converge 
  weak$^*$ (as $N\to \infty$) to $\rho\,  d\mathcal{H}_d$.

  Furthermore, if $\inf_A\rho>0$ and $\rho$ is upper semi-continuous, then any 
  $(w,s)$-energy minimizing sequence of configurations $\omega_N$ is well-separated in the sense
  of Theorem 4 with $\alpha =d$.  
\end{corollary}

{\bf Remark:} The first part of Corollary \ref{wrho} holds for $s=d$ when $A$ is contained in a $C^1$ $d$-dimensional manifold.

\bigskip

 Finally, we consider weight functions with isolated zeros.   
  For $t > 0$, we say that a   function $w:A\times A\to \RR$ has a {\it   zero at $(a,a)\in D(A)$ of order at most $t$} if 
 there are positive constants $C$ and $\delta$ such that 
 \begin{equation}
w(x,y)\ge C|x-a|^t   \qquad  ( x,y  \in A \cap B_{d'}(a,\delta)).
\end{equation}
  If $w$ has a zero $a\in A$ whose order  is too large, then $a$ may act as an attractive ``sink'' with  $\mathcal E_s^w(A,N)=0$.  For  example, let 
 $A$ be the closed unit ball in $\RR^d$, $w(x,y)=|x|^t+|y|^t$ for $x,y\in A$ with  $t>s>d$.  If $\omega_N=\{x_1,\ldots, x_N)$ is a configuration of $N$ points in $A$, then   
 $E_s^w(\gamma \omega_N)=\gamma^{t-s}E_s^w(\omega_N)$ for any $0<\gamma<1$.
 Taking $\gamma \to 0$, shows that $\mathcal E_s^w(A,N)=0$.

 A closed set $A\subset \RR^{d'}$ is {\em $\alpha$-regular at $a\in A$
 } if there  are positive constants $C_0$ and $\delta$ such that 
 \begin{equation}\label{zero0}
  (C_0)^{-1}r^\alpha\le \mathcal{H}_\alpha(A\cap B_{d'}(x,r))\le C_0 r^\alpha
 \end{equation}             
 for all $x\in A\cap B_{d'}(a,\delta)$ and $0<r<\delta$.   
 
 \begin{theorem}\label{zerothm}
Let $A\subset \RR^{d'}$ be a compact $d$-rectifiable set and $s>d$.  Suppose  
  $A$ is $\alpha_i$-regular with $\alpha_i \le d$ at $a_i$, $i=1,\ldots, n$,  for a finite collection of points $a_1, \ldots, a_n$    in $A$ and that  $w:A\times A\to [0,\infty]$  is    a CPD-weight function
  on $K\times K$ for any compact    $K\subset A\setminus \{a_1,\ldots, a_n\}$. If  
   $w$ has a zero of order  at most $t<s$ at each $(a_i,a_i)$, 
  then the conclusions of Theorem~\ref{rth1} hold. 
 \end{theorem}

  {\bf Remark:}  The hypotheses of Theorem~\ref{zerothm} imply that 
 $\int_{A}{(w(x,x))^{-d/s}d\mathcal H_d(x)}<\infty$ (see Section 6).

\section{Lemmas.}

In this  section  we prove several lemmas which are central to the proofs of our main theorems.   

\subsection{Divide and conquer.}

In this subsection we provide two lemmas relating the minimal energy problem on $A=B\cup D$ to the  minimal energy problems on $B$ and $D$, respectively.

In order to unify our computations for the cases $s>d$ and $s=d$,  we define, for  integers $N>1$,
$$
\tau_{s,d}(N):=\begin{cases} N^{1+s/d}, &  s >d,\\
N^2\log N, & s =d   \end{cases}
$$
and set $\tau_{s,d}(N)=1$ for  $N=0$ or $1$. For a set $A\subset
\RR^{d'}$ and $s\ge d$, let
$$
\UL g^w_{s,d}(A):=\liminf_{N\to \infty}{\frac {\mathcal
E^w_s(A,N)}{\tau _{s,d}(N)}}, \qquad \OL g^w_{s,d}(A):=
\limsup_{N\to \infty}{\frac{\mathcal E^w_s(A,N)}{\tau _{s,d}(N)}},
$$
and
$$
g^w_{s,d}(A):=\lim_{N\to \infty}{\frac{\mathcal E^w_s(A,N)}{\tau
_{s,d}(N)}}$$ if this limit exists (these quantities are allowed
to be infinite). In the case $w(x,y)\equiv 1$, we     use the notations $\UL
g_{s,d}(A)$, $\OL g_{s,d}(A)$ and $g_{s,d}(A)$, respectively.

Let ${\rm dist}(B,D):=\inf{\{\left|x-y\right| : x\in B,\ y\in
D\}}$ denote the  distance between sets $B,D\subset\RR^{d'}$. The
following two lemmas extend  Lemmas 3.2  and
3.3 from \cite{HarSaf04} to the weighted case.   We remark that the following
results hold  when quantities are 0 or infinite  using  $0^{-d/s}=0^{-s/d}=\infty$ and
$\infty ^{-d/s}=\infty ^{-s/d}=0$.
\begin {lemma}\label {rl2}
Let $s \geq d>0$ and suppose that $B$ and $D$ are sets in
$\RR^{d'}$ such that ${\rm dist}(B,D)>0$.   Suppose $w:(B\cup D)\times (B\cup D)\to[0,\infty]$ is
bounded on the subset $B\times D$. Then
\begin {equation}\label {r2.1}
\OL g^w_{s,d}(B\cup D)^{-d/s}\geq \OL g^w_{s,d}(B)^{-d/s}+\OL
g^w_{s,d}(D)^{-d/s}.
\end{equation}
\end{lemma}
\begin{proof}
Assume that $0<\OL
g^w_{s,d}(B),\ \OL g^w_{s,d}(D)<\infty$. Denote
$$
\alpha^\ast:=\frac {\OL g^w_{s,d}(D)^{d/s}}{\OL
g^w_{s,d}(B)^{d/s}+\OL g^w_{s,d}(D)^{d/s}}.
$$
For $N\in \NN$, let $N_B:=\lfloor\alpha^\ast N\rfloor$ (where $\lfloor x \rfloor$ denotes
the  greatest integer less than or equal to $x$), $N_D:=N-N_B$
and $\omega ^B_N\subset B$ and $\omega^D_N\subset D$ be
configurations of $N_B$ and $N_D$ points respectively such that $
E^w_s(\omega^B_N)<\mathcal E^w_s(B,N_B)+1$ and
$E^w_s(\omega^D_N)<\mathcal E^w_s(D,N_D)+1$. Let
$\gamma_0:={\rm dist}(B,D)>0$. Then
\begin {align*}
\mathcal E^w_{s,d}(B\cup D,N)&\leq E^w_s(\omega^B_N\cup
\omega^D_N)
\\
&=E^w_s(\omega^B_N)+E^w_s(\omega^D_N)+2\!\!\!\sum_{x\in\omega^B_N
,\ y\in\omega^D_N}{\frac{w(x,y)}{\left|x-y\right|^s}}
\\
&\leq \mathcal E^w_s(B,N_B)+\mathcal
E^w_s(D,N_D)+2+2\gamma_0^{-s}N^2\|w\|_{B\times D },
\end{align*}
where $\|w\|_{B\times D }$ denotes the supremum of $w$ over $B\times D$.
Dividing by $\tau_{s,d}(N)$ and taking into account that
$\tau_{s,d}(N_B)/\tau_{s,d}(N)\to (\alpha^\ast)^{1+s/d}$ as $N\to
\infty$, we obtain
\begin {align*}
\OL g^w_{s,d}(B\cup D)&\leq \limsup_{N\to\infty}{\frac {\mathcal
E^w_s(B,N_B)}{\tau_{s,d}(N)}}+\limsup_{N\to\infty}{\frac {\mathcal
E^w_s(D,N_D)}{\tau_{s,d}(N)}}
\\
&=\limsup_{N\to\infty}{\frac {\mathcal E^w_s(B,N_B)}{
\tau_{s,d}(N_B)}\cdot \frac {\tau_{s,d}(N_B)}{
\tau_{s,d}(N)}}+\limsup_{N\to\infty}{\frac {\mathcal
E^w_s(D,N_D)}{\tau_{s,d}(N)}}
\\
&\leq \OL g^w_{s,d}(B)\cdot (\alpha^\ast)^{1+s/d}+\OL
g^w_{s,d}(D)\cdot (1-\alpha^\ast)^{1+s/d}
\\
&=\(\OL g^w_{s,d}(B)^{-d/s}+\OL g^w_{s,d}(D)^{-d/s}\)^{-s/d}.
\end{align*}
The remaining cases when $\OL g^w_{s,d}(B)$ or $\OL g^w_{s,d}(D)$ are 0 or $\infty$ easily follow from the monotonicity of $\OL g^w_{s,d}$.
\end{proof}

The following statement in particular shows sub-additivity of $\UL g^w_{s,d}(\cdot)^{-d/s}$.

\begin {lemma}\label {rl3}
Let $s\geq d>0$ and $B,D\subset\RR ^{d'}$. Suppose  $w:(B\cup D)\times (B\cup D)\to[0,\infty]$.
Then
\begin {equation}\label {r3.1}
\UL g^w_{s,d}(B\cup D)^{-d/s}\leq \UL g^w_{s,d}(B)^{-d/s}+\UL
g^w_{s,d}(D)^{-d/s}.
\end{equation}
Furthermore, if $\UL g^w_{s,d}(B),\UL g^w_{s,d}(D)>0$ and
at least one of these quantities is finite,
then
\begin {equation}\label {number}
\lim_{\mathcal N\ni N\to\infty}{\frac {\left|\W\omega_N\cap
B\right|}{N}}=\frac {\UL g^w_{s,d}(D)^{d/s}}{\UL
g^w_{s,d}(B)^{d/s}+\UL g^w_{s,d}(D)^{d/s}} 
\end{equation}
holds for any sequence $\{\W\omega_N\}_{N\in \mathcal N}$ of
$N$-point configurations in $B\cup D$ such  that
\begin {equation}\label {w}
\lim_{\mathcal N\ni N\to\infty}{\frac
{E^w_s(\W\omega_N)}{\tau_{s,d}(N)}}=\(\UL g^w_{s,d}(B)^{-d/s}+\UL
g^w_{s,d}(D)^{-d/s}\)^{-s/d},
\end {equation}
where  $\mathcal N$ is some infinite subset of
$\NN$.
\end{lemma}
In the case $\UL g^w_{s,d}(D)=\infty$ the right-hand side of
relation (\ref {number}) is understood to be 1.

\begin{proof}
Assume that $0<\UL g^w_{s,d}(B),\UL
g^w_{s,d}(D)<\infty$ (we leave other cases to the reader). Let an
infinite subset $\mathcal N_1\subset \NN$ and a sequence of
point configurations $\{\omega _N\}_{N\in \mathcal N_1}$, $\omega_N\subset B\cup D$, be such that $
\lim_{\mathcal N_1\ni N\to\infty}{{\left|\omega_N\cap
B\right|}/{N}}=\alpha $, where $0\le \alpha\le 1$.  Set $N_B:=\left|\omega_N\cap B\right|$ and $N_D:=\left|\omega_N\setminus B\right|$. Then
$$
E^w_{s}(\omega_N)\geq E^w_{s}(\omega_N\cap
B)+E^w_s(\omega_N\setminus B)\geq \mathcal E^w_{s}(B,N_B)+\mathcal
E^w_s(D,N_D),
$$
and we have
\begin {align*}
\liminf_{\mathcal N_1\ni N\to\infty}{\frac
{E^w_s(\omega_N)}{\tau_{s,d}(N)}}&\geq \liminf_{\mathcal N_1\ni
N\to\infty}{\frac {\mathcal E^w_{s}(B,N_B)}{\tau_{s,d}(N_B)}\cdot
\frac {\tau_{s,d}(N_B)}{\tau_{s,d}(N)}}
\\
&\qquad +\liminf_{\mathcal N_1\ni N\to\infty}{\frac {\mathcal
E^w_{s}(D,N_D)}{\tau_{s,d}(N_D)}\cdot \frac
{\tau_{s,d}(N_D)}{\tau_{s,d}(N)}}
\end{align*}
\begin {equation}\label {qq}
\ \ \ \ \ \ \ \ \ \ \ \ \ \ \ \ \ \ \ \ \ \ \ \ \ \ \ \ \ \ \ \ \ \ \ \ \geq F(\alpha):=\UL g^w_{s,d}(B)\alpha ^{1+s/d}+\UL
g^w_{s,d}(D)(1-\alpha)^{1+s/d}.
\end{equation}
Let 
$$
\W\alpha:=\frac {\UL g^w_{s,d}(D)^{d/s}}{\UL g^w_{s,d}(B)^{d/s}+\UL
g^w_{s,d}(D)^{d/s}},
$$
and $\{\W\omega_N\}_{N\in \mathcal N}$ be any sequence of point
sets satisfying (\ref {w}). If $\mathcal N_2\subset \mathcal N$ is
any infinite subsequence such that the quantity
$\left|\W\omega_N\cap B\right|/N$ has a limit as $\mathcal N_2\ni
N\to \infty$ (denote it by $\alpha _1$), then in the case $\UL
g^w_{s,d}(B),\UL g^w_{s,d}(D)<\infty$ by (\ref {w}) and (\ref
{qq}) we have
$$
F(\W\alpha)=\lim_{\mathcal N_2\ni N\to \infty}{\frac
{E^w_s(\W\omega_N)}{\tau_{s,d}(N)}}\geq F(\alpha_1).
$$
It is not difficult to see that $\W\alpha$ is the only minimum
point of $F(t)$ on $[0,1]$. Hence $\alpha_1=\W\alpha$, which proves
(\ref {number}).


Now let  $\{\OL\omega_N\}_{N\in \mathcal N_3}$ be a sequence of
$N$-point configurations in $B\cup D$ such that
$$
\UL g^w_{s,d}(B\cup D)=\lim_{\mathcal N_3\ni N\to \infty}{\frac
{E^w_s(\OL \omega_N)}{\tau_{s,d}(N)}}
$$
($\OL\omega _N$'s can be chosen for example so that $E^w_s(\OL
\omega_N)<\mathcal E^w_s(B\cup D,N)+1$). If $\mathcal N_4\subset
\mathcal N_3$ is such an infinite set that $\lim_{\mathcal N_4\ni
N\to \infty}{\left|\OL\omega_N\cap B\right|/N}$ exists (denote it
by $\alpha_2$), then by (\ref {qq}) we obtain
\begin {align*}
\UL g^w_{s,d}(B\cup D)&=\lim_{\mathcal N_4\ni N\to \infty}{\frac
{E^w_s(\OL\omega_N)}{\tau_{s,d}(N)}}\geq F(\alpha_2)
\\
&\geq F(\W\alpha)=\(\UL g^w_{s,d}(B)^{-d/s}+\UL
g^w_{s,d}(D)^{-d/s}\)^{-s/d},
\end{align*}
which implies (\ref{r3.1}).
\end{proof}

\subsection{Lemmas from geometric measure theory}
Recall that $\beta_d$ denotes the  volume of the unit ball in $\RR^d$.  For convenience, we also define $\beta_0:=1$.   For a set $W\subset \RR^{d'}$ and $h>0$, we let
$$
W(h):=\{x\in \RR^{d'} : {\rm dist}(x,W)<h\}.
$$
The upper and the lower Minkowski contents of the set $W$ are defined, respectively, by
$$
\OL {\mathcal M}_d(W):=\limsup_{r\to 0^+}{\frac {\mathcal L_{d'}[W(r)]}{\beta_{d' -d}\cdot r^{d' -d}}} \text{ \, and }
\ \ \ \UL {\mathcal M}_{d}(W):=\liminf_{r\to 0^+}{\frac {\mathcal L_{d'}[W(r)]}{\beta_{d' -d}
\cdot r^{d' -d}}}.
$$
 If the upper and the lower Minkowski contents of the set $W$ coincide, then
this common value, denoted  by $\mathcal M_{d}(W)$,  is called
{\em the  Minkowski content of $W$}.
We shall rely on the following property of closed $d$-rectifiable sets.
\begin {lemma}(see {\rm \cite[Theorem 3.2.39]{FedGMT}}).\label {rl6}
If $W\subset \RR^{d'}$ is a closed $d$-rectifiable set, then $\mathcal M_{d}(W)=\mathcal H_d(W)$.
\end {lemma}
We shall also need the following fundamental lemma from geometric measure theory.
\begin {lemma}(see \cite [Lemma 3.2.18]{FedGMT}).\label {rl1}
Let $W\subset \RR^{d'}$ be a $d$-rectifiable set.
Then for every $\epsilon >0$ there exist compact sets $K_1$,
$K_2$, $K_3,\ldots \subset \RR^d$ and bi-Lipschitz mappings
$\psi_i:K_i\to \RR^{d'}$ with constant $1+\epsilon$,
$i=1,2,3,\ldots$, such that $\psi_1(K_1)$, $\psi_2(K_2)$,
$\psi_3(K_3), \ldots$ are disjoint subsets of $W$ with
$$
\mathcal H_d\(W\setminus\bigcup \limits_{i}{\psi_i(K_i)}\)=0.
$$
\end{lemma}

In fact, the above lemma holds for any set of finite $\mc{H}_d$-measure that is, up to a set of $\mc{H}_d$-measure zero,
the countable union of $d$-rectifiable sets.  However, Lemma~\ref{rl6} does not hold for this larger class (cf.  \cite{FedGMT}).\\

\subsection{Regularity Lemma}

To get an estimate from below for $\UL g_{s,d}(A)$ we will need the following result.
\begin {lemma}\label {rl5}
Let $s>d$ and  suppose $A\subset \RR^{d'}$ is a
compact set such that ${\mathcal M}_d(A)$ exists and is finite. Then for every $\epsilon \in (0,1)$ there is some
$\delta
>0$ such that for any compact set $K\subset A$ with $\UL {\mathcal
M}_{d}(K)>\mathcal M_{d}(A)-\delta$ we have
\begin{equation}\label{reg}
\UL g_{s,d}(A)\geq (1-\epsilon)\UL g_{s,d}(K).
\end {equation}
\end{lemma}
\begin{proof}  The
assertion of the lemma holds trivially  if $\UL g_{s,d}(A)=\infty$. Hence, we   assume
$\UL g_{s,d}(A)<\infty$. Let $\mathcal
N\subset \NN$ be an infinite subset such that
$$
\lim_{\mathcal N \ni N\to \infty}{\frac {\mathcal
E_s(A,N)}{N^{1+s/d}}}=\UL g_{s,d}(A).
$$
Choose $\delta \in\(0,1/2^{4d}\)$ and
set
\begin{equation}\label{rhohN}
\rho :=\delta ^{1/(4d)} \quad
\text{ and } \quad
 h_N:=\frac {1}{3}\rho ^2N^{-1/d},\qquad N\in \mathcal N.
\end{equation}
Suppose $K$ is a compact subset of $A$
such that $\UL {\mathcal M}_{ d}(K)>\mathcal M_{d}(A)-\delta$.  Then there is some
$N_\delta\in \NN$ such that for any $N>N_\delta$, $N\in \mathcal
N$, we have
\begin {equation}\label {r5.1}
\frac {\mathcal L_{d'}[A(h_N)]}{\beta _{d'
-d}h_N^{d' -d}}\leq \mathcal M_{d}(A)+\delta\ \ \ {\rm and
}\ \ \ \frac {\mathcal L_{d'}[K(h_N)]}{\beta _{d'
-d}h_N^{d' -d}}\geq \mathcal M_{d}(A)-\delta .
\end {equation}
For  $N\in \mathcal N$ with $N>N_\delta$, let
$\omega^\ast_N:=\{x_{1,N},\ldots,x_{N,N}\}$  be an  $s$-energy minimizing $N$-point configuration on $A$.
For $i=1,\ldots, N$,  let  $r_i^N:=\min\limits_{j\neq
i}{\left|x_{j,N}-x_{i,N}\right|}$  denote the distance from $x_{i,N}$ to its nearest neighbor  in $\omega^\ast_N$.
Further, we partition $\omega^\ast_N$ into  a ``well-separated'' subset  $$
\omega ^1_N:=\{x_{i,N}\in \omega^\ast_N : r^N_i\geq \rho
N^{-1/d}\},
$$
and its complement $\W \omega ^1_N:=\omega^\ast_N\setminus \omega
^1_N$.  We next show that $\omega ^1_N$ has sufficiently many
points. (In the case $\mathcal H_d(A)>0$, Theorem \ref {th2}
implies that $\left|\omega^1_N\right|=N$ for sufficiently small
$\rho$. However, the case $\mathcal H_d(A)=0$ still requires
consideration and we provide a proof   that works in either
case). For $N\in \mathcal N$, we obtain
\begin{align*}
\mathcal
E_s(A,N)&=E_s(\omega^\ast_N)=\sum\limits_{i=1}^{N}{\sum\limits_{j=1\atop
j\neq i}^{N}{\frac {1}{\left|x_{i,N}-x_{j,N}\right|^s}}}\geq
\sum\limits_{i=1}^{N}{\frac {1}{(r^N_i)^s}}
\\
&\geq \sum\limits_{x_{i,N}\in \W \omega ^1_N}{\frac
{1}{(r^N_i)^s}}\geq \sum\limits_{x_{i,N}\in \W \omega ^1_N}{\frac
{1}{\(\rho N^{-1/d}\)^s}}= \left|\W \omega ^1_N\right|\rho
^{-s}N^{s/d}.
\end{align*}
Let $k_0:=\UL g_{s,d}(A)+1$. There is $N_1\in \NN$ such that for
any $N>N_1$, $N\in \mathcal N$,
$$
\frac {\mathcal E_s(A,N)}{N^{1+s/d}}<k_0.
$$
For the rest of the proof of this lemma, let $N\in \mathcal N$ be greater than $N_2:=\max
\{N_1,N_\delta\}$. Then,
$$
\frac {\left|\W \omega ^1_N\right|}{\rho ^{s}N}\leq \frac
{\mathcal E_s(A,N)}{N^{1+s/d}}<k_0,$$
and, hence, we have
\begin{equation}
\label{omega1bnd}\left|\W \omega
^1_N\right|<k_0\rho ^sN \text{ and }\left|\omega ^1_N\right|>(1-k_0\rho
^s)N.
\end{equation}

Next we consider
$$
\omega ^2_N:=\omega ^1_N\bigcap K(3h_N),\ \ \ \W \omega
^2_N:=\omega ^1_N\setminus K(3h_N), 
$$
and show that the cardinality of $\omega ^2_N$ is sufficiently large. From  (\ref {r5.1}) we
get
\begin{align}
\!\mathcal L_{d'}[A(h_N)\setminus &K(h_N)] =\mathcal
L_{d'}[A(h_N)]-\mathcal L_{d'}[K(h_N)]\label {r10}
\\
&\leq (\mathcal M_{d}(A)+\delta)\beta _{d' -d}h_N^{d'
-d}   -(\mathcal
M_{d}(A)-\delta)\beta _{d' -d}h_N^{d' -d}\nonumber \\
&=2\beta _{d' -d}\delta
h_N^{d' -d}. \nonumber
\end{align}
Note, that
\begin{equation}\label{FN}
F_N:=\bigcup_{x\in \W \omega ^2_N}{B_{d'}(x,h_N)}\subset
A(h_N)\setminus K(h_N).
\end{equation}

For any distinct points $x_{i,N},x_{j,N}\in \W \omega ^2_N$  we have
$$\left|x_{i,N}-x_{j,N}\right|\geq r^N_i\geq \rho N^{-1/d}>\rho
^2N^{-1/d}=3h_N.$$
   Hence, $B_{d'}(x_{i,N},h_N)\bigcap
B_{d'}(x_{j,N},h_N)=\emptyset$. Then, using (\ref {r10}) and (\ref{FN}), we
get
\begin {align*}
\left|\W \omega ^2_N\right|&=
\left(\beta
_{d'}h_N^{d'}\right)^{-1}\sum\limits_{x\in \W \omega ^2_N}{\mathcal
L_{d'}[B_{d'}(x,h_N)]}=\left(\beta
_{d'}h_N^{d'}\right)^{-1}\mathcal L_{d'}(F_N)\\
&\leq\left(\beta
_{d'}h_N^{d'}\right)^{-1} \mathcal L_{d'}[A(h_N)\setminus K(h_N)]\leq
2\beta _{d' -d}\beta
_{d'}^{-1}\delta h_N^{-d}.
 \end{align*}
Hence, recalling (\ref{rhohN}),  we have
\begin {align}\label{omega2bnd}
\left|\W \omega ^2_N\right|&\leq 2\cdot 3^d\beta _{d' -d}\beta _{d'}^{-1}\delta ^{1/2}N.
\end {align}
Let $\chi_0:=2\cdot 3^d\beta _{d' -d}\beta _{d'}^{-1}$.
Then, using (\ref{omega1bnd}) and (\ref{omega2bnd}),  we have
$$
\left|\omega ^2_N\right|=\left|\omega ^1_N\right|-\left|\W \omega
^2_N\right|\geq \(1-k_0\rho ^{s}-\chi_0\delta ^{1/2}\)N.
$$

Next, we choose a configuration $\omega ^K_N$ of points in $K$ which is close to $\omega ^2_N$ and has the same number of points and order of the minimal $s$-energy as  $\omega ^2_N$. For every
$x_{i,N}\in \omega ^2_N$ pick a point $y_{i,N}\in K$ such that
$\left|x_{i,N}-y_{i,N}\right|<3h_N=\rho ^2N^{-1/d}$ and let
$\omega ^K_N:=\{y_{i,N} : x_{i,N}\in \omega ^2_N\}$. Since every
point $x_{i,N}\in \omega ^2_N$ lies in $\omega ^1_N$, we
have
$$
\left|x_{i,N}-y_{i,N}\right|<\rho ^2N^{-1/d}\leq\rho r^N_i\leq
\rho \left|x_{i,N}-x_{j,N}\right|, \ \ j\neq i.
$$
Then, if $x_{i,N}\neq x_{j,N}$ are points from $\omega ^2_N$, we
have
\begin {align*}
\left|y_{i,N}-y_{j,N}\right|&=\left|y_{i,N}-x_{i,N}+x_{i,N}-x_{j,N}+x_{j,N}-y_{j,N}\right|
\\
&\geq\left|x_{i,N}-x_{j,N}\right|-\left|x_{i,N}-y_{i,N}\right|
-\left|x_{j,N}-y_{j,N}\right|
\\
&\geq\left|x_{i,N}-x_{j,N}\right|-2\rho
\left|x_{i,N}-x_{j,N}\right|=(1-2\rho)\left|x_{i,N}-x_{j,N}\right|.
\end {align*}
Since $\rho\in(0,1/2)$, it follows that $\left|\omega ^K_N\right|=\left|\omega ^2_N\right|$
and
\begin {align*}
E_s(\omega^\ast_N)&=\sum\limits_{x\neq y\in \omega^\ast_N}{\frac
{1}{\left|x-y\right|^s}}\geq \sum\limits_{x\neq y\in \omega
^2_N}{\frac {1}{\left|x-y\right|^s}}
\\
&\geq (1-2\rho)^s\sum\limits _{x\neq y\in\omega ^K_N}{\frac
{1}{\left|x-y\right|^s}}=(1-2\rho)^sE_s(\omega ^K_N).
\end {align*}

Now suppose $\epsilon\in(0,1)$. We may choose $\delta>0$  sufficiently small (recall $\rho=\delta ^{1/(4d)}$)
so that   $(1-2\rho)^s(1-k_0\rho ^{s}-\chi_0\delta^{1/2})^{1+s/d}\ge (1-\epsilon)$.
Hence,
\begin {align*}
\UL g_{s,d}(A)&=\lim_{\mathcal N\ni N\to
\infty}{\frac{E_s(\omega^\ast_N)}{ N^{1+s/d}}}\geq (1-2\rho)^s
\liminf_{\mathcal N\ni N\to \infty}{\frac{E_s(\omega ^K_N)}{
N^{1+s/d}}}
\\
&\geq (1-2\rho)^s \liminf_{\mathcal N\ni N\to \infty}{\frac
{\mathcal E_s(K,\left|\omega ^2_N\right|)}{\left|\omega
^2_N\right|^{1+s/d}} \cdot \(\frac {\left|\omega
^2_N\right|}{N}\)^{1+s/d}}
\\
&\geq (1-2\rho)^s\(1-k_0\rho ^{s}-\chi_0\delta
^{1/2}\)^{1+s/d}  \liminf_{N\to \infty}{\frac{\mathcal
E_s(K,N)}{ N^{1+s/d}}}\\ &\geq (1-\epsilon)\UL g_{s,d}(K) 
\end{align*}
holds for any   compact subset $K\subset A$
such that $\UL {\mathcal M}_{ d}(K)>\mathcal M_{d}(A)-\delta$. \end{proof}

\section{Proofs of  Main Theorems.}



\begin{proof}[Proof of Theorem~\ref{rectifiable}]

First we remark that if $K\subset \RR^d$ is compact, then $K$ is trivially a $d$-rectifiable manifold (or set) and so Theorem A shows, for  $s>d$,
\begin {equation}\label{wA}
g_{s,d}(K)=\frac
{C_{s,d}}{\mathcal L_d(K)^{s/d}}.
\end{equation}

Suppose $0<\epsilon<1$.   Since $A\subset\RR^{d'}$ is a compact $d$-rectifiable set,
  Lemma \ref {rl1} implies the  existence of compact sets $K_1$, $K_2$, $K_3$,
$\ldots\subset \RR^{d}$ and bi-Lipschitz mappings $\psi_i:K_i\to
\RR^{d'}$, $i=1,2,3,\ldots$, with constant $1+\epsilon$ such
that $\psi_1(K_1)$, $\psi_2(K_2)$, $\psi_3(K_3),\ldots$ are
disjoint subsets of $A$ whose union covers $\mathcal H_d$-almost
all of $A$.

Let $n$ be large enough so that
$${\mathcal H}_d\left(\bigcup_{i=1}^n \psi_i(K_i)\right)=\sum_{i=1}^n  {\mathcal H}_d(\psi_i(K_i)) \ge (1+\epsilon)^{-d}\, {\mathcal H}_d(A).$$  Since $\psi_i$ is bi-Lipschitz with constant $(1+\epsilon)$ we have
\begin{align}\label{psi}
\OL g_{s,d}(\psi_i(K_i))&\le (1+\epsilon)^s  g_{s,d}(K_i)=C_{s,d} (1+\epsilon)^s {\mathcal L}_d(K_i)^{-s/d}
\\ &\le C_{s,d} (1+\epsilon)^{2s} {\mathcal H}_d(\psi_i(K_i))^{-s/d}. \nonumber
\end{align}
Applying Lemma~\ref{rl2} we obtain
\begin{align}\label{gupbnd}
\OL g_{s,d}&(A)\le \OL g_{s,d}\left(\bigcup_{i=1}^n \psi_i(K_i)\right)\le
\left(\sum_{i=1}^n \OL g_{s,d}(\psi_i(K_i))^{-d/s}\right)^{-s/d}\\
&\le C_{s,d} (1+\epsilon)^{2s}
\left(\sum_{i=1}^n  {\mathcal H}_d(\psi_i(K_i))\right)^{-s/d} \le C_{s,d} (1+\epsilon)^{3s} {\mathcal H}_d(A)^{-s/d}. \nonumber
\end{align}

We next provide a lower bound for $\UL g_{s,d}(A)$.
Since $A$ is a closed $d$-rectifiable set,  we have  $\mathcal M_d(A)=\mathcal H_d(A)<\infty$
(cf. Lemma \ref {rl6}). Let $\delta>0$ be as in   Lemma
\ref {rl5}, i.e.,  inequality (\ref{reg}) holds for every compact set $K\subset A$
such that $ {\mathcal M}_{ d}(K)>\mathcal M_{d}(A)-\delta$.   Now let  $n'$ be large enough so that
$$
 {\mathcal M}_{ d}\left(\cup _{i=1}^{n'}{\psi _i(K_i)}\right)=\sum\limits_{i=1}^{n'}{\mathcal H_d[\psi _i(K_i)]}>\mathcal H_d(A)-\delta =\mathcal M_{d}(A)-\delta.
$$
As in (\ref{psi})  we have
\begin{align} \label{psi2}
\UL g_{s,d}(\psi_i(K_i))&\ge (1+\epsilon)^{-s}  g_{s,d}(K_i)=C_{s,d} (1+\epsilon)^{-s} {\mathcal L}_d(K_i)^{-s/d}
\\ &\ge C_{s,d} (1+\epsilon)^{-2s} {\mathcal H}_d(\psi_i(K_i))^{-s/d}.\nonumber
\end{align}

  Then    Lemma
\ref {rl3} and (\ref {psi2}) give
\begin {align} \label{glowbnd}
\UL g_{s,d}(A)&\ge (1-\epsilon)\UL g_{s,d}\(\cup _{i=1}^{n'}{\psi _i(K_i)}\)\ge
(1-\epsilon)\left(\sum\limits_{i=1}^{n'}{\UL g_{s,d}[\psi
_i(K_i)]^{-d/s}}\right)^{-s/d}
 \\
&\geq \frac{(1-\epsilon)C_{s,d}}{(1+\epsilon)^{2s}}\left(\sum\limits_{i=1}^{n'}{\mathcal H_d[\psi_i(K_i)]}\right)^{-s/d}\geq \frac{(1-\epsilon)C_{s,d}}{(1+\epsilon)^{2s}}\mathcal H_d(A)^{-s/d}. \nonumber
\end{align}
Letting $\epsilon$ go to zero in (\ref{gupbnd}) and (\ref{glowbnd}), we obtain (\ref{w3.1}).
 \end{proof}



\subsection{Proofs of Theorems \ref{rth1} and \ref{wth1H}}
The following lemma relates   the weighted minimal energy problem ($s\ge d$) on a set $A\subset\RR^{d'}$ to the unweighted minimal energy problem on   compact subsets of $A$.   Theorems \ref{rth1} and \ref{wth1H}
then follow easily from this lemma.

For convenience, we denote $$C_{d,d}:=\beta_d, \ \ d\in \NN. $$
\begin {lemma}\label{wth1}
Suppose $s\geq d$, $A\subset \RR^{d'}$ is compact with
 $\mathcal H_d (A)<\infty$, and that $w$  is a CPD-weight function   on $A\times A$.
 Furthermore, suppose that  for
any compact subset $K\subset A$,
 the limit $g_{s,d}(K)$ exists and is given by
\begin {equation}\label {w*1}
g_{s,d}(K)=
\frac{C_{s,d}}{\mathcal H_d (K)^{s/d}}.
\end {equation}
Then
\begin{enumerate}
\item[(a)] $g^w_{s,d}(A)$ exists and is given by
\begin {equation}\label {www1}
g^w_{s,d}(A)= C_{s,d}\(\mathcal
H^{s,w}_{d}(A)\)^{-s/d},
\end{equation}
and,
\item[(b)] if a sequence
$\{\W\omega_N\}_{N=2}^\infty$, where
$\W\omega_N=\{x^N_1,\ldots,x^N_N\}$, is asymptotically
($w,s$)-energy minimizing on the set $A$ and $\mathcal H_d(A)>0$, then
\begin {equation}\label{www2}
\frac 1N\sum\limits_{k=1}^{N}{\delta _{x^N_k}}{\ \ \ast \over}\!\!\!\to h^{s,w}_d,\ \ N\to \infty.
\end{equation}
\end{enumerate}
\end{lemma}
{\bf Remark.} If $\mathcal H_d (K)=0$, condition (\ref {w*1}) is
understood as $g_{s,d}(K)=\infty$.

\begin{proof} To prove the first part of the theorem, we
break $A$ into disjoint ``pieces" of small diameter and estimate the ($w,s$)-energy of $A$ by
replacing $w$ with its
supremum or infinum on each of the ``pieces" and applying Lemmas \ref {rl2} and \ref {rl3}.

 For $\delta>0$,
suppose that $\mathcal P_\delta$ is a partition of $A$ such that
 ${\rm diam\, }P\leq \delta$ and $\mathcal H_d(\OL P)=\mathcal H_d( P)$ for $P\in \mathcal{P}_\delta$,
 where
 $\OL B$ denotes the closure of a set $B$.
For each $P\in \mathcal P_\delta$, choose a closed subset $Q_P\subset P$ so that
 $\mathcal Q_\delta:=\{Q_P: P\in \mathcal P_\delta\}$ satisfies
 \begin {equation}\label {w03}
\sum\limits_{P\in \mathcal P_\delta}{\mathcal H_d(Q_P)}\geq \mathcal H_d(A)-\delta .
\end{equation}

An example of such systems $\mathcal{P}_\delta$ and $\mathcal{Q}_\delta$ can be constructed as follows.
Let $G_j[t]$ be the hyperplane in $\RR^{d'}$ consisting of all points whose $j$-th coordinate equals $t$. If $(-a,a)^{d'}$ is a cube containing $A$, then for
${\bf i}=(i_1,\ldots,i_{d'})\in \{1,\ldots, m\}^{d'}$, let
$$
R_{\bf i}:=[t^1_{i_1-1},t^1_{i_1})\times\cdots\times[t^{d'}_{i_{d'}-1},t^{d'}_{i_{d'}}),
$$
where $m$ and partitions $-a=t^j_0<t^1_1<\ldots<t^j_m=a$,  $j=1,\ldots, d'$, are chosen so that the diameter of every $R_{\bf i}$, ${\bf i}\in\{1,\ldots, m\}^{d'}$, is less than $\delta$ and $\mathcal H_d(G_j[t^j_i]\cap A)=0$ for all $i$ and $j$. (Since $\mathcal H_d(A)<\infty$, there are at most countably many values of $t$ such that $\mathcal H_d(G_j[t]\cap A)>0$.) Then,  we may choose $$\mathcal{P}_\delta=\{R_{\bf i}\cap A : {\bf i}\in\{1,\ldots, m\}^{d'}\}$$   and    $\gamma\in(0,1)$
  sufficiently close to 1 such that   (\ref {w03}) holds for $\mathcal{Q}_\delta=\{Q_{\bf i}: {\bf i}\in\{1,\ldots, m\}^{d'}\}$,  where $Q_{\bf i}=\left(\gamma(\OL R_{\bf i}-c_{\bf i})+c_{\bf i}\right)\cap A$ and $c_{\bf i}$ denotes the center of $R_{\bf i}$.

For $B\subset A$, let
$$
\OL w_{B}=\sup_{x,y\in B}{w(x,y)}
\text{ and }
\UL w_{B}=\inf_{x,y\in B}{w(x,y)}
$$
and define the simple functions   $\OL w_\delta(x):=\sum_{P\in \mathcal P_\delta}{\OL w_{P} \cdot\chi_{P}(x)}$
and $\UL w_\delta(x):=\sum_{P\in \mathcal P_\delta}{\UL w_{P} \cdot\chi_{P}(x)}$,   
where ${\chi }_{_K}$ denotes the characteristic function of a set $K$.
Since the distance between any two sets from $\mathcal Q_\delta$ is strictly positive,   Lemma \ref{rl2} and equation (\ref {w*1}) imply
\begin {eqnarray}\label{w01}
&\OL g^w_{s,d}(A)^{-d/s}\geq \OL g^w_{s,d} \left(\bigcup_{Q\in \mathcal{Q}_\delta}Q \right)^{-d/s}\geq \sum\limits_{Q\in \mathcal Q_\delta\atop Q\neq \emptyset}{\(\OL w_{Q}\cdot \OL g_{s,d}(Q)\)^{-d/s}}
\\ \nonumber
&=C_{s,d}^{-d/s}\sum\limits_{Q\in \mathcal{Q}_\delta\atop Q\neq \emptyset}{\OL w^{-d/s}_{Q}\cdot \mathcal H_d(Q)}\geq C_{s,d}^{-d/s}\int\limits_{\bigcup Q \atop {Q\in \mathcal{Q}_\delta}}({\OL w_\delta (x))^{-d/s}d\mathcal H_d(x)}.
\end{eqnarray}
   Applying Lemma \ref {rl3} and relation (\ref {w*1}), we similarly have
\begin {eqnarray}\label {w02}
&\UL g^w_{s,d}(A)^{-d/s}\leq \sum\limits_{P\in \mathcal P_\delta}{\(\UL w_{P}\cdot \UL g_{s,d}(P)\)^{-d/s}}= \sum\limits_{P\in \mathcal P_\delta}{\(\UL w_{P}\cdot \UL g_{s,d}(\OL P)\)^{-d/s}}
\\ \nonumber
&=C_{s,d}^{-d/s}\sum\limits_{P\in \mathcal P_\delta}{\UL w_{P}^{-d/s}\cdot\mathcal H_d(\OL P)}=
C_{s,d}^{-d/s}\int\limits_{A}({\UL w_\delta (x))^{-d/s}d\mathcal H_d(x)}.
\end {eqnarray}
Since $w$ is a CPD-weight  function on $A\times A$, there is some neighborhood $G$ of $D(A)$ such that
$\eta:=\inf_G w>0$.    For $\delta>0$ sufficiently small, we have $P\times P\subset G$ for
all $P\in \mathcal{P}_\delta$, and hence
$\OL w_\delta(x) \geq w(x,x) \geq \UL w_\delta(x)\ge \eta$ for $x\in A$.  Furthermore, $w$ is continuous at  $(x,x)\in D(A)$ for $\mathcal H_d$-almost all $x\in A$ and  thus, for any such $x$, it follows that  $\OL w_\delta(x) $ and $\UL w_\delta(x)$ converge to
$w(x,x)$ as $\delta\to 0$.    Therefore,   by the Lebesgue Dominated Convergence Theorem,
the integrals $$\int\limits_{\bigcup Q \atop {Q\in \mathcal{Q}_\delta}}{\(\OL w_\delta (x)\)^{-d/s}d\mathcal H_d(x)} \text{ and  } \int_{A}{\(\UL w_\delta (x)\)^{-d/s}d\mathcal H_d(x)}$$  both converge to $\mathcal H_d^{s,w}(A)$ as $\delta\to 0$.  Hence, using  (\ref {w01}) and (\ref {w02}),
we obtain (\ref {www1}).

Now suppose that $\mathcal H_d(A)>0$ and $\W\omega_N=\{x^N_1,\ldots,x^N_N\}$ is an asymptotically
($w,s$)-energy minimizing sequence of $N$-point configurations on $A$. It is well-known \cite[p.~9]{LanFMPT}
 that
the weak$^*$ convergence result given in (\ref {www2}) is equivalent to the
assertion that
\begin {equation}\label{**}
\lim\limits_{N\to \infty}{\frac {\left|\W\omega_N\bigcap
B\right|}{N}}=h^{s,w}_{d} (B)
\end{equation}
holds for any almost clopen subset $B\subset A$, where we recall that a set $B\subset A$ is called {\it almost clopen (with respect to $A$ and $\mathcal H_d$)}, if the
$\mathcal H_d$-measure of the relative boundary of $B$ with
respect to $A$ equals zero.

If $B\subset A$ is almost clopen, then the hypotheses of Lemma \ref {wth1} imply
%
%
\begin{align*}
\lim_{N\to\infty}{\frac {E^w_s(\W\omega_N)}{\tau_{s,d}(N)}}&=C_{s,d}\(\mathcal H_d^{s,w}(A)\)^{-s/d}\\ &=C_{s,d}\(\mathcal H^{s,w}_d(\OL B)+\mathcal H^{s,w}_d(\OL {A\setminus B})\)^{-s/d}\\
&=\(g^w_{s,d}(\OL B)^{-d/s}+g^w_{s,d}(\OL {A\setminus B})^{-d/s}\)^{-s/d}.
\end{align*}
Using relation (\ref {number}) in Lemma \ref {rl3} and (\ref {www1}) for $\OL B$ and $\OL{A\setminus B}$, we get
$$
\lim\limits_{N\to \infty}{\frac {\left|\W\omega_N\bigcap
B\right|}{N}}=\frac { g^w_{s,d}(\OL {A\setminus B})^{d/s}}{
g^w_{s,d}(\OL B)^{d/s}+ g^w_{s,d}(\OL {A\setminus B})^{d/s}}=h^{s,w}_d(B)
$$
showing that (\ref {www2}) holds.
\end{proof}

Theorems \ref {rth1} and \ref {wth1H} then follow from Lemma~\ref{wth1} and
Theorems~\ref{rectifiable} and ~\ref{thmB} as we now explain.
If $s>d$ and $A\subset \RR^{d'}$ is a closed $d$-rectifiable set, then
every compact subset $B\subset A$ is also closed and $d$-rectifiable
and Theorem~\ref{rectifiable} implies that $B$ satisfies condition
(\ref {w*1})  and so Theorem \ref {rth1} then follows from Lemma~\ref{wth1}. If $s=d$ and $A$ is a compact subset of a
$d$-dimensional $C^1$-manifold in $\RR^{d'}$, then applying Theorem~\ref{thmB}   to every compact subset of $A$, we get (\ref {w*1}).
Consequently Theorem \ref{wth1H} follows from  Lemma \ref {wth1} with $s =d$.

\section {Proofs of the separation results: Theorem~ \ref {th2} and Corollary~\ref{cor1}.}

In this subsection we prove Theorem~ \ref {th2} and Corollary~\ref{cor1}. For the proof of
these results we will need Frostman's lemma establishing the existence of a non-trivial measure on $A$
satisfying a regularity assumption similar to the one in
\cite{MarMayRahSaf02} for arclength.
\begin {lemma}\label {Frostman}(see e.g. \cite [Theorem 8.8]{MatGSMES}).
Let $\alpha>0$ and $A$ be a Borel set in $\RR^{d'}$. Then
$\mathcal H_\alpha(A)>0$ if and only if there is a Radon measure
$\mu$ on $\RR^{d'}$ with compact support contained in $A$ such
that $0<\mu(A)<\infty$ and \beq \label {regu} \mu
\[B_{d'}(x,r)\]\leq r^\alpha,\ \ \ x\in \RR^{d'},\ \ r>0. \eeq
Moreover, one can find $\mu$ so that $\mu (A)\geq
c_{d',\alpha}\mathcal H^\infty_\alpha (A)$, where $c_{d',\alpha}>0$
is independent of $A$.
\end {lemma}

We proceed using the technique developed in \cite{KuiSaf98}. Let
$\omega^\ast_N:=\{x_1,\ldots,x_N\}$, $N\in \NN$, $N\geq 2$, be a
$(w,s)$-energy minimizing configuration on $A$ (for convenience,
we dropped the subscript $N$ in writing energy minimizing points
$x_{k,N}$). For $i=1,\ldots,N$ let
$$
U_i(x):=\sum_{j\neq i}{\frac {w(x,x_j)}{\left|x-x_j\right|^s}}, \
\ x\in A.
$$
From the minimization property we have that $U_i(x_i)\leq U_i(x)$, $x\in
A$, $i=1,\ldots,N$. If $\mu$ is a measure from Lemma \ref
{Frostman}, set
$
r_0:=\( {\mu (A)}/{2N}\)^{1/\alpha}
$
and let
$$
D_i:=A\setminus \bigcup \limits_{j\neq i}{B_{d'}(x_j,r_0)},\ \
i=1,\ldots,N.
$$
Then, by the properties of $\mu$, we have
$$
\mu (D_i)\geq \mu (A)-\sum_{j\neq i}{\mu \[B_{d'}(x_j,r_0)\]}\geq
\mu (A)-(N-1)r_0^\alpha > \frac {\mu (A)}{2}>0,
$$
$i=1,\ldots,N$. Consequently, 
\begin {align*}
U_i(x_i)&\leq \frac {1}{\mu (D_i)}\int\limits_{D_i}{U_i(x)d\mu
(x)}\leq \frac {2}{\mu (A)}\sum_{j\neq i}{\int\limits_{D_i}{\frac
{w(x,x_j)}{\left|x-x_j\right|^s}d\mu (x)}}
\nonumber\\
&\leq\frac {2\|w\|}{\mu (A)}\sum_{j\neq i}{\int\limits_{A\setminus
B_{d'}(x_j,r_0)}{\frac {1}{\left|x-x_j\right|^s}d\mu (x)}},\ \
i=1,\ldots,N,\nonumber
\end{align*}
where $\|w\|:=\sup \{\left|w(x,y)\right| : x,y\in A\}$. Let $R:={\rm
diam}\ A$. Then by (\ref {regu}) we have $\mu (A)\leq R^\alpha$.
For every $y\in A$ and $r\in (0,R]$, using (\ref {regu}) we also
get
\begin{align*}
T_s(y,r)&:=\int\limits_{A\setminus B_{d'}(y,r)}{\frac
{1}{\left|x-y\right|^s}d\mu (x)}=\int\limits_{0}^{r^{-s}}{\mu
\{x\in A : \frac {1}{\left|x-y\right|^s}>t\}dt}\\
&=\frac
{\mu(A)}{R^{s}}
+ \int\limits_{R^{-s}}^{r^{-s}}{\mu
\[B_{d'}\(y,t^{-1/s}\)\]dt}
\leq
R^{\alpha-s}+\int\limits_{R^{-s}}^{r^{-s}}{t^{-\alpha/s}dt}
\\
&\leq
\begin{cases} \frac {s}{(s-\alpha)}r^{\alpha-s},& s>\alpha,\\
1+\alpha\ln \frac{R}{r}, & s=\alpha. \end{cases}
\end{align*}
Then for $i=1,\ldots, N$ and $s>\alpha$ we have
\begin {equation}\label {w15}
U_i(x_i)\leq \frac {2\|w\|}{\mu (A)}\sum\limits_{j\neq
i}{T_s(x_j,r_0)}\leq\frac {2s(N-1)\|w\|}{(s-\alpha)\mu
(A)r_0^{s-\alpha}}\leq C_1\|w\|\(\frac {N}{\mu (A)}\)^{s/\alpha},
\end {equation}
where $C_1>0$ is a constant independent of $A$, $w$ and $N$. Hence,
$$
\mathcal
E^w_s(A,N)=E^{w}_s(\omega^\ast_N)=\sum\limits_{i=1}^{N}{U_i(x_i)}\leq
\frac {M_{s,\alpha}\|w\|}{\mathcal H^\infty_\alpha
(A)^{s/\alpha}}N^{1+s/\alpha},
$$
where $M_{s,\alpha}$ is a constant independent of $A$, $w$, and
$N$. In particular, when $w\equiv 1$, we get
$$
\mathcal E_s(A,N)\leq\frac
{s2^{s/\alpha}N^{1+s/\alpha}}{(s-\alpha)(c_{d',\alpha})^{s/\alpha}\mathcal H^\infty_\alpha
(A)^{s/\alpha}}.
$$
Since $w$ is a CPD-weight function, there are $\eta,\rho>0$ such
that $w(x,y)>\eta$ whenever $\left|x-y\right|<\rho$. Assume that
$\delta (\omega ^\ast_N)<\rho$ and let $i_s$ and $j_s$ be such
that $\delta (\omega ^\ast_N)=\left|x_{i_s}-x_{j_s}\right|$. Then
with some constant $C_2>0$ independent of $N$ and the choice of
$\omega^\ast_N$ we obtain from (\ref {w15})
$$
C_2N^{s/\alpha}\geq U_{i_s}(x_{i_s})\geq \frac
{w(x_{i_s},x_{j_s})}{\left|x_{i_s}-x_{j_s}\right|^s}\geq \frac
{\eta}{\left|x_{i_s}-x_{j_s}\right|^s}=\frac {\eta}{\delta
(\omega^\ast_N)^s}.
$$
Hence,
$$
\delta (\omega ^\ast_N)\geq C_0N^{-1/\alpha},
$$
where $C_0=C_0(A,w,\alpha,s)>0$. Thus, in any case, $$\delta
(\omega ^\ast_N)\geq \min \{\rho, C_0 N^{-1/\alpha}\}\geq
C_sN^{-1/\alpha}, \ \ N\geq 2,$$ for a sufficiently  small constant
$C_s>0$ independent of $N$ and $\omega^\ast_N$. In particular,
when $w\equiv 1$, we have
$$
\delta (\omega^\ast_N)\geq \frac {c_{s,\alpha}}{(\mathcal
H^\infty_\alpha (A)\cdot N)^{1/\alpha}}.
$$
The case $s=\alpha$ is handled analogously, which  completes the proofs of Theorem \ref {th2} and  Corollary~\ref{cor1}.

\section{Proof of Theorem~\ref{zerothm}}
The essential ingredient in the proof of Theorem~\ref{zerothm} is the following lemma which
assumes lower regularity.
We say that a set $K\subset \RR^{d'}$ is {\em lower $\alpha$-regular} if there are
  positive constants $C_0$ and $r_0$ so that 
  \begin{equation}\label{lowerreg}
  (C_0)^{-1}r^\alpha \le \mathcal{H}_\alpha(K\cap B_{d'}(x,r))
 \end{equation}  
 for all $x\in K$ and $r<r_0$.
 \begin{lemma}\label{zerolemma}
Suppose $K\subset\RR^{d'}$ is compact and lower $\alpha$-regular and   $a\in K$.   Further suppose 
 $s>\alpha$ and $w:K\times K\to [0,\infty]$   is a  CPD-weight function on $K'\times K'$ for any compact   $K'\subset K\setminus \{a\}$.   If $w$  
 has a zero of order at most $t$ at $(a,a)$, where  
   $0<t<s$,  then  
   \begin{equation}\label{zerolemmaIneq}
   \UL g^w_{s,\alpha}(K)\ge C_1 C_0^{-s/\alpha} 2^{-(s+t)} \left(\int_{K}\frac{1}{|x-a|^{(t\alpha)/s}}\, d\mathcal H_\alpha(x)\right)^{-s/\alpha}.
   \end{equation}
 \end{lemma}
 \begin{proof}
 Let $\omega_N=\{x_1,\ldots, x_N\}$ be a configuration of $N$ distinct points in $K$.  For  $i=1,\ldots, N$, let $\rho_i=|x_i-a|$,   $r_i=\min_{j\neq i}|x_i-x_j|$, and choose $y_i\in \omega_N$  such that $|x_i-y_i|=r_i$.   Since $K$ is bounded,   there is some finite $L$  (independent of $N$)
 such that there are at most $L-1$ of the points $x_i\in \omega_N$ with the property  that $r_i\ge r_0$.  We  order the 
 points in $\omega_N$ so that $\rho_N\le \rho_i$ for  $i=1,\ldots, N$ and so that $r_i< r_0$ for $i=1,\ldots, N-L$.    
 It follows from Cauchy's and Jensen's inequality (see (29) of \cite{HarSaf04}) that 
 if $\gamma_1, \ldots, \gamma_M$  are positive numbers, then
 \begin{equation}
 \sum_{i=1}^M\gamma_i^{-s}\ge M^{1+s/\alpha}\left(\sum_{i=1}^M\gamma_i^\alpha\right)^{-s/\alpha} 
 \end{equation}
from which we obtain
 \begin{align} \label{zero1}
E_s^w(\omega_N)&\ge \sum_{i=1}^{N-L} \frac{w(x_i,y_i)}{r_i^s}\ge  C_1\sum_{i=1}^{N-L} \frac{\rho_i^t}{r_i^s} \\
&\ge  C_1(N-L)^{1+s/\alpha}\left(\sum_{i=1}^{N-L}\frac{r_i^\alpha}{\rho_i^{t\alpha/s}}\right)^{-s/\alpha}. \nonumber
\end{align}
For $i=1,\ldots, N-1$, observe that $$r_i=\min_{j\neq i}|x_i-x_j|\le |x_i-a|+\min_{j\neq i}|a-x_j|\le \rho_i+\rho_N\le 2\rho_i $$
and so 
\begin{equation} \label{zero2}
|x-a|\le |x-x_i|+|x_i-a|\le r_i/2+\rho_i\le 2\rho_i,  \qquad (x\in B(x_i,r_i/2)).
\end{equation} 
 
Using (\ref{lowerreg}) and (\ref{zero2}) we have
$$
\frac{r_i^\alpha}{\rho_i^{t\alpha/s}} \le C_0  2^{(\alpha/s)(s+t)}\int_{K\cap B(x_i,r_i/2)}\frac{1}{|x-a|^{t\alpha/s}}\, d\mathcal H_\alpha(x) 
$$
for $i=1,\ldots,N-L$. 
Since  $ B(x_i,r_i/2)$ and $ B(x_j,r_j/2)$ are disjoint for $i\neq j$, it follows that 
$$
\sum_{i=1}^{N-L}\frac{r_i^\alpha}{\rho_i^{t\alpha/s}}\le C_0 2^{(\alpha/s)(s+t)}\int_{K}\frac{1}{|x-a|^{t\alpha/s}}\, d\mathcal H_\alpha(x),
$$
which combined with (\ref{zero1}) completes the proof. 
 \end{proof}
 
 {\bf Remark:}
If $K$ is $\alpha$-regular at $a$ in the above lemma, then the integral 
$\int_{K}\frac{1}{|x-a|^{(t\alpha)/s}}\, d\mathcal H_\alpha(x)$
appearing in (\ref{zerolemmaIneq}) is finite  (cf. \cite[p. 109]{MatGSMES})
and thus the Lebesgue Dominated Convergence Theorem gives
$$\lim_{\delta\to 0}  \int_{K\cap B_{d'}(a,\delta)}\frac{1}{|x-a|^{(t\alpha)/s}}\, d\mathcal H_\alpha(x)=0 $$  and so
$\lim_{\delta\to 0} g^w_{s,\alpha}(K\cap B_{d'}(a,\delta))=\infty.$

\bigskip

Now we are prepared to complete the proof of Theorem~\ref{zerothm}.  
First note that the hypotheses of Theorem~\ref{zerothm} (namely that  $A$ is   $\alpha_i$-regular  at $a_i$ and $w$ has a zero of order of at most $t<s$ at $a_i$ for $i=1,\ldots, n$) imply that 
$\int_{A}w(x,x)^{-s/d}\, d\mathcal{H}_d(x)\le \infty.$

Suppose $\epsilon>0$ .  By  Lemma~\ref{zerolemma} and 
  Lemma~\ref{rl3}  we can  find $\delta>0$ such that 
$B_\epsilon:=\bigcup_{i=1}^n (A\cap B_{d'}(a_i,\delta))$ satisfies $\UL g^w_{s,d}(B_\epsilon)\ge \epsilon^{-1}$   (note that if $\alpha<d$ and $\UL g^w_{s,\alpha}(K)>0$, then $\UL g^w_{s,d}(K)=\infty$) and
$\mathcal{H}_d^{s,w}(A_\epsilon) = \int_{A_\epsilon}w(x,x)^{-s/d}\, d\mathcal{H}_d(x)\ge (1-\epsilon)\mathcal{H}_d^{s,w}(A)$, where $A_\epsilon:=A\setminus B_\epsilon$.

Since $w$ is a CPD-weight function on $A_\epsilon\times A_\epsilon$,
  it follows from Theorem~\ref{rth1} that $g^w_{s,d}(A_\epsilon)$ exists and equals 
$C_{s,d}\mathcal{H}_d^{s,w}(A_\epsilon)^{-s/d}$.   Lemma~\ref{rl3} then gives
\begin{align}\label{zerothm1}
\UL  g^w_{s,d}(A)&\ge (  g^w_{s,d}(A_\epsilon)^{-d/s}+\UL g^w_{s,d}(B_\epsilon)^{-d/s})^{-s/d}\\
&\ge  (C_{s,d}^{-d/s}\mathcal{H}_d^{s,w}(A_\epsilon)+ \epsilon^{d/s})^{-s/d} \nonumber \\
&\ge  (C_{s,d}^{-d/s}\mathcal{H}_d^{s,w}(A)+ \epsilon^{d/s})^{-s/d}. \nonumber
\end{align}
Also, we clearly have 
\begin{equation}\label{zerothm2}\OL g^w_{s,d}(A)\le g^w_{s,d}(A_\epsilon)=C_{s,d}\mathcal{H}_d^{s,w}(A_\epsilon)^{-s/d}\le C_{s,d}(1-\epsilon)^{-s/d}\mathcal{H}_d^{s,w}(A)^{-s/d}.  
\end{equation}

Taking $\epsilon\to 0$ in (\ref{zerothm1}) and (\ref{zerothm2}) shows 
that $g^w_{s,d}(A)$ exists and equals 
$C_{s,d}\mathcal{H}_d^{s,w}(A )^{-s/d}$.    If $\mathcal{H}_d^{s,w}(A ) >0$, then, as in the proof of Theorem~\ref{rth1},
Lemma~\ref{rl3} implies that (\ref{wdistlim}) holds for any asymptotically $(w,s)$-energy minimizing sequence of configurations
$\W\omega_N=\{x^N_1,\ldots,x^N_N\}$, $N=2,3,\ldots$,
 for $A$ which completes the proof of Theorem~\ref{zerothm}.

\end {document}